\documentstyle[aps,prb,psfig]{revtex}
\begin{document} 
\title {The Structure, Dynamics and Electronic Structure of \\
Liquid Ag-Se Alloys Investigated by Ab Initio Simulation}
\author{F. Kirchhoff, J.  M.  Holender and M.  J.  Gillan} 
\address{ Physics Department, Keele University \\ 
Keele, Staffordshire ST5 5BG, U.K.}
\date{\today} 
\maketitle 
\begin{abstract}
{\em Ab initio} molecular-dynamics simulations have been used to
investigate the structure, dynamics and electronic properties of
the liquid alloy Ag$_{1-x}$Se$_x$ at 1350~K and at the three
compositions $x =$~0.33, 0.42 and 0.65. To provide a point of
reference, calculations are also presented for the equilibrium structure
and the electronic structure of the $\alpha$-Ag$_2$Se crystal. The
calculations are based on density-functional theory in the local
density approximation and on the pseudopotential plane-wave method.
For the solid, we find excellent agreement with experiment for
the equilibrium lattice parameters and the atomic coordinates of the
12-atom orthorhombic unit cell, and we present an analysis of the electronic
density of states and density distribution. The reliability
of the liquid simulations is confirmed by detailed comparisons with very
recent neutron diffraction results for the partial structure
factors and radial distribution functions (RDF) of the stoichiometric
liquid Ag$_2$Se. Comparison with the predictions of an empirical
interaction model due to Rino {\em et al.} is also given for $\ell$-Ag$_2$Se.
The {\em ab initio} simulations show a dramatic change of the Se-Se
RDF with increasing Se content. This change is due to the formation
of Se clusters bound by covalent bonds, the Se-Se bond length being
almost the same as in pure $c$-Se and $\ell$-Se. The clusters are 
predominantly chain-like, but for higher $x$ there is a significant
fraction of 3-fold coordinated Se atoms. It is shown that the 
equilibrium fractions of Se present as isolated atoms and in clusters
can be understood on a simple charge-balance model based on an ionic
interpretation. The Ag diffusion coefficient in the simulated stoichiometric
liquid is consistent with experimental values measured in the
high-temperature superionic solid. The Ag and Se diffusion coefficients
both increase with Se content, in spite of the Se clustering. An analysis
of the Se-Se bond dynamics reveals surprisingly short bond lifetimes
of less than 1~ps. The electronic density of states (DOS) for 
$\ell$-Ag$_2$Se strongly resembles that of the solid. Some of 
the changes of DOS with composition arise directly from the
formation of Se-Se covalent bonds. Results for the electronic
conductivity $\sigma$ obtained using the Kubo-Greenwood approximation
are in adequate agreement with experiment for $\ell$-Ag$_2$Se, but 
for the high Se contents the simulation results for $\sigma$ are 3--4
times greater than experimental values. Possible reasons for this
are discussed.
\end{abstract}
\twocolumn
\section{Introduction}
The study of liquid binary alloys has been an extremely fruitful source
of insights into the relations between structure and electronic
properties in condensed matter.
There has been a vast amount of experimental work
on metallic and semiconducting binary liquids, and it has been well
known for many years that their properties often vary dramatically
with composition\cite{end90,hensel79}. 
Famous cases are the Cs-Au and Mg-Bi alloys\cite{schmuzler76,enderby70}, 
in which
the pure elements are good metals, but nevertheless the stoichiometric
mixtures (CsAu and Mg$_3$Bi$_2$) have very low conductivities, and
their structures are characteristic of molten salts. These effects arise from
the electronegativity difference between the elements, and the resulting
charge transfer, partial ionicity and heterocoordination. In systems
where one of the elements is a semiconductor in the liquid state --
for example alloys of metals with S or Se -- even richer behavior
can be expected, since variation of composition should change the
bonding from metallic through partially ionic to covalent. We report
here a set of simulations of the liquid Ag-Se system performed
using {\em ab initio} molecular dynamics (AIMD), which we have
used to explore these effects.

The electrical conductivity $\sigma$ of the Ag$_{1-x}$Se$_x$ 
system has been measured
over most of the composition range at temperatures from 973
to 1573~K\cite{endo80,glazov86,ohno94,ohno96}.
As the composition goes from pure $\ell$-Ag to the stoichiometric
alloy Ag$_2$Se, $\sigma$ decreases from the typically metallic value
of $\sim$~50000~$\Omega^{-1}$cm$^{-1}$ to typical semiconducting 
values of $\sim$~500~$\Omega^{-1}$cm$^{-1}$. This strongly 
suggests the formation of a 
pseudogap in the electronic density of states at the stoichiometric
composition, with the Fermi energy lying in this gap. As the
Se content is further increased, $\sigma$ 
varies rather little over the range between $x=0.33$ and $x=0.65$, and
then descends to very low values of $\sim$~1~$\Omega^{-1}$cm$^{-1}$ at
700~K as the Ag content goes to zero\cite{sigma}. The properties of pure $\ell$-Se have been
very extensively studied\cite{tamura90}, and it is known to have a gap of
{\em ca.}~1.9~eV in the density of states just above its melting point
($\sim$490~K), so that it is a wide-gap semiconductor. This gap decreases
with increasing temperature, and appears to be {\em ca.}~0.7~eV at 1350~K and
100~bar~\cite{hosokawa90}.

Pure $\ell$-Ag has the rather closed-packed structure and high
coordination number ({\em ca.}~12) expected from its f.c.c. crystal
structure\cite{waseda}. Until very recently, the structure of $\ell$-Ag$_2$Se
had been measured only at the level of the neutron-weighted
structure factor\cite{price93}, but neutron diffraction combined with isotope
variation has now been used to measure the partial structure
factors and hence the partial radial distribution functions\cite{barnes96}. 
These reveal the heterocoordination characteristic of ionic or partially
ionic liquids, as will be discussed more fully below.
Diffraction studies on $\ell$-Se have been made by several independent
groups\cite{bellissent80,edeling81,tamura92}, and it is well established 
that the average coordination of the
atoms is close to 2.0 over a wide range of temperatures and pressures
from the melting point up to 1773~K and 815~bar. It is widely believed that 
this coordination indicates the presence of extended chain-like structures, 
and there is support for this from simulation work\cite{hohl91,selam}. These 
facts imply that there must
be major changes of atomic ordering as the composition is varied
from pure $\ell$-Ag to pure $\ell$-Se. However, nothing is yet
known about the structure of the alloys except at the  stoichiometric
composition.

The aim of the work reported here is to use {\em ab initio} molecular
dynamics simulation (AIMD) \cite{cp85,df} to investigate the structure, 
the atomic dynamics and the electronic structure of the liquid Ag-Se alloys at
different compositions. The issues we want to address are: the
relationship between the electronic structure of the liquid
and the solid; the characteristics of the electronic density of
states and electron distribution as a function of composition;
the variation of liquid structure with composition, and particularly
how the apparently ionic structure of $\ell$-Ag-Se goes over to
the chain-like structure of $\ell$-Se; and the way this liquid structure
is related to the dynamics of the ions. At present, there is almost no
experimental information about any of these questions. Of course,
before we address any of these issues, we have to show that our
AIMD techniques are able to give a faithful representation of the real
material. Because of this, we will begin by presenting calculations
on the $\alpha$-Ag$_2$Se crystal, which has a rather complex structure
and gives a good test of our methods. Our study of the crystal also
provides an essential reference point for discussing the properties
of the liquid.

AIMD is ideally suited to this type of problem. The energetics of the
system and the forces on the ions are calculated by first-principles
quantum methods, with no adjustable parameters, and without the need for
the empirical interionic potentials used in earlier simulations
on many liquid mixtures, including $\ell$-Ag$_2$Se \cite{rino88}. In addition, the 
thermal-equilibrium structure, atomic dynamics and electronic
structure are all calculated together in a completely unified and
self-consistent way. AIMD techniques have been used to investigate
a number of other liquid metals and semiconductors, including the binary
alloys K-Si\cite{galli91}, Na-Sn\cite{schone95}, Cs-Pb\cite{wijs94} 
and Ga-Se\cite{gase96}. Preliminary results of the work presented
here have already been published elsewhere \cite{letter}.
The rest of the paper is organised as follows. In Sec.\ II, we summarize
the simulation techniques, and we then present in Sec.\ III our calculations
on the equilibrium structure and the electronic structure of $\alpha$-Ag$_2$Se.
Our simulation results on the liquid Ag-Se alloys are reported in Sec.\ IV, 
where we examine successively the 
structure, the atomic dynamics and the electronic properties, including
the density of states and the electronic conductivity. 
We present a brief, discussion of the relation between atomic ordering
and electronic properties in Sec.\ V, and our conclusions are
summarized in Sec.\ VI.

\section{Computational methods}

Our {\em ab initio} molecular dynamics (AIMD) technique is closely
related to the Car-Parrinello approach \cite{cp85,df}. We use density 
functional theory within the local density 
approximation (LDA)~\cite{hohenberg64,kohn65,jones89}. 
Only the valence electrons are treated explicitly, the interaction of
the valence electrons with the atomic cores being represented by
norm-conserving non-local pseudopotentials\cite{hsc,bhs}.  
The simulations are performed in 
a periodically repeated cell with the wave functions expanded in plane 
waves. The basis set includes all plane waves whose kinetic energy
is less than a chosen cut-off energy $E_{\rm cut}$.

An important difference between our calculations and the original
Car-Parrinello approach is that instead of treating the electronic
degrees of freedom by `fake dynamics' we relax the electrons to the
Born-Oppenheimer surface for every ionic configuration, using
conjugate-gradients minimization \cite{gil89,sti89cg,pay92}. As in the original
scheme, we calculate the forces acting on the ions
via the Hellmann-Feynman theorem, and these are then used to integrate
the classical equation of motion of the ions.  It is computationally
more expensive to bring the electrons to the Born-Oppenheimer surface
than to do one step in the Car-Parrinello method, but the extra cost is
compensated by the fact that we can use a larger molecular dynamics
time step. To handle the semi-metallic nature of the system, we use 
Fermi-surface smearing, with the electronic occupation numbers treated 
as auxiliary dynamical variables \cite{gil89,kre94,gru94}. A detailed
report on how this is done in practice has been presented in our work
on $\ell$-Ga\cite{ga95}.

Although AIMD has become a standard technique, very few simulations
have been reported on systems containing transition or post-transition
metals \cite{pas92,kre93}. Because of the important role of the
$d$-electrons, it is essential to include them explicitly, and this
brings two kinds of problems. Firstly, the number of occupied orbitals
is substantially increased. Secondly, the compact nature of the
$d$-orbitals means that a large cutoff $E_{\rm cut}$ is needed, and
this implies a large plane-wave basis set. 
Our AIMD simulations on the Ag-Se system were performed with a version
of the CETEP code\cite{cetep} running on a Cray T3D parallel
supercomputer.
We note that the version of the code used here
reaches the ground state by minimizing with respect to all bands
simultaneously, rather than by the band-by-band method of earlier
versions.  In the liquid simulations, we needed about 12 minimization 
steps for every ionic configuration to converge the total free energy 
within 10$^{-5}$~eV/atom.

Technical details of our {\em ab initio} pseudopotentials are as
follows.  The Ag pseudopotential has been optimised using the method of
Lin {\em et al.} \cite{lin}, which is a refinement of the scheme due to
Rappe {\em et al.} \cite{rap}.  The Se pseudopotential does not require
optimisation and the standard Kerker method \cite{ker} suffices.  For
Ag, all states below $4d$, $5s$ and $5p$  are treated as core states.
The $s$ and $p$ components of the Ag pseudopotential were generated
using the atomic configuration 4$d^{10}$5$s^{0.25}$5$p^{0.25}$, and the
$d$ component using the configuration 4$d^{10}$5$s^{0.5}$. The core
radii were 2.0, 2.0 and 2.5 a.u. for the $s$, $p$ and $d$ components
respectively.  For Se, we used 4$s^{2}$4$p^{4}$ for the $s$ and $p$
waves and 4$s^{2}$4$p^{2.75}$4$d^{0.25}$ for the $d$ wave; the core
radii were chosen to be 2.0, 2.0, and 2.3 a.u. for the $s$, $p$ and $d$
components  respectively.  We work with the pseudopotentials in
Kleinman-Bylander separable form\cite{kb}, with the $s$-wave treated as
local; the non-local parts of the pseudopotentials are treated in real
space \cite{kin91}.  We use a plane-wave cut-off of 400~eV.
As a preliminary test of our pseudopotentials, we have
performed calculations on the low temperature phase $\alpha$-Ag$_2$Se,
which we present in more detail in the next section. We also tested
the pseudopotentials in our work on AgCl\cite{agcl} and  the liquid
Ga-Se alloy\cite{gase96}.

\section{Solid ${\rm A\lowercase{g}}_2{\rm S\lowercase{e}}$}

\label{sec:solid}
At atmospheric pressure silver selenide, Ag$_2$Se, exists in two 
polymorphic forms. The high temperature form, $\beta$-Ag$_2$Se, is
stable above 406~K. The structure is b.c.c. with spacing
$a = 4.983$~\AA, and the unit cell contains two Ag$_2$Se units. The
Se atoms form a stable b.c.c lattice, while the Ag atoms are statistically
distributed among several interstitial sites\cite{rahl36,gmelinag}.
In this structure, all the Ag atoms are mobile, and
in fact $\alpha$-Ag$_2$Se is a typical superionic conductor similar to
$\alpha$-AgI and $\alpha$-Ag$_2$S \cite{saku77,cava77,cava80}.

\setlength{\unitlength}{1mm}
\begin{figure}[b]
\begin{picture}(85,100)(0,0)
\psfig{figure=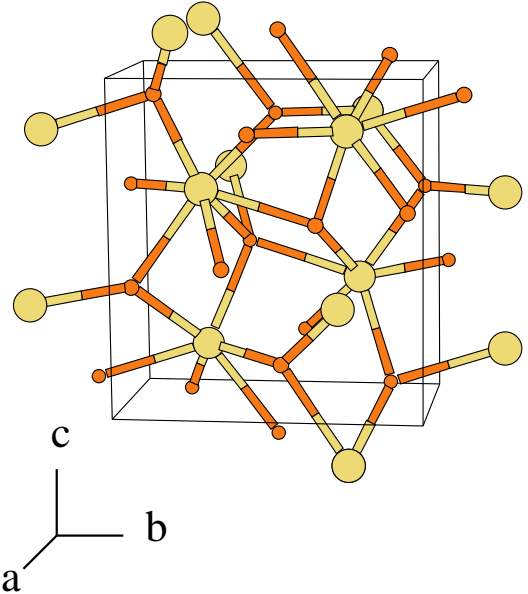,width=8.5cm}
\end{picture}
\caption{Arrangement of Ag and Se atoms in solid \protect{$\alpha$-Ag$_2$Se.} 
Ag and Se atoms are represented by small dark spheres and large light spheres,
respectively.}
\label{fig:solid}
\end{figure}
Ag$_2$Se transforms from the $\beta$ phase to a low temperature form,
$\alpha$-Ag$_2$Se, at 406~K. The structure of $\alpha$-Ag$_2$Se has been
resolved by X-ray powder diffraction as orthorhombic with cell
constants $a=4.333$, $b=7.062$ and $c=7.764$~\AA. The space group is
P2$_1$2$_1$2$_1$ and there are four Ag$_2$Se units in the unit
cell\cite{wieg71}. A three dimensional representation of the structure
is shown is in Fig.\ \ref{fig:solid}. The four Se atoms are equivalent
by symmetry, and they almost lie on two planes perpendicular to the $b$
axis. There are two inequivalent Ag sites. The first type,
Ag$_{\mbox{\scriptsize I}}$, lie just above or below the planes of
the chalcogen atoms. These atoms are tetrahedrally coordinated by Se
atoms, three in the nearest Se plane and one in the next, at distance
2.62, 2.79, 2.86 and 2.71~\AA, respectively. The other silver atoms,
Ag$_{\mbox{\scriptsize II}}$, lie almost halfway between the planes of
chalcogen atoms. The arrangement of the Se around the
Ag$_{\mbox{\scriptsize II}}$ atoms is almost triangular, at distances
2.72, 2.74 and 2.81~\AA, with two second neighbors at 3.28 and
3.50~\AA. Each Se atom is surrounded by seven Ag atoms,
or nine if the second neighbors are taken into account.

Silver is well known to exist in various solid-state compounds with
coordination numbers ranging from 2 up to 8\cite{wells}.  Examples of
coordination 2, 3, and 4 are given by Ag$_2$O\cite{wyckoff},
Ag$_3$AsS$_3$\cite{engel68} and AgI\cite{wells}.
In this respect the binary compounds of silver with the chalcogen 
elements, Ag$_2$X (X=O, S, Se and Te), are remarkable since the coordination 
of silver varies from linear to trigonal bipyramidal when going down the 
VIa column\cite{wieg71}. With a valence of one for silver these compounds
may be regarded as ionic which makes these low coordinations all the
more striking, since ionic bonds usually imply higher coordination.
The stability of the Ag $d^{10}$ ion in such compounds may be attributed
to the mixing of the silver $d$ and $s$ states which occurs when the atom
is placed in a low-symmetry environment\cite{orgel58,burd92}.

The electronic properties of $\alpha$-Ag$_2$Se are those of a semiconductor
with a small energy gap. The exact value of the energy gap, is however 
controversial, and values ranging from 0.02 to 0.22~eV have been
published\cite{glazov86,gmelinag}.

\begin{table}
\caption{Comparison of calculated and experimental crystallographic parameters
for \protect{$\alpha$-Ag$_2$Se} (experimental values in parentheses). 
The lattice constants are given in \AA~.
The internal parameters $x$, $y$, $z$ are the coordinates of the atoms in units 
of $a$, $b$ and $c$.  The three other positions equivalent by symmetry are 
given by ($\frac{1}{2}$+$x$,$\bar{y}$,$\bar{z}$), 
($\bar{x}$,$\frac{1}{2}-y$,$\frac{1}{2}$+$z$) and 
($\frac{1}{2}-x$,$\frac{1}{2}$+$y$,$\frac{1}{2}-z$).}
\setdec 0.0
\label{tab:solid}
\begin{tabular}{lccc}
Lattice constants	&  $a$	&  $b$	&  $c$  \\
\tableline
		    &  4.218  &  6.949  &  7.649 \\
		    & (4.333) & (7.062) & (7.764) \\
\tableline
Internal parameters      	& $x$	&  $y$	&  $z$  \\
\tableline
	  Ag$_{\rm{I}}$   &  0.109  &  0.367  &  0.455  \\
	        & (0.107) & (0.366) & (0.456) \\
	  Ag$_{\rm{II}}$  &  0.729  &  0.027  &  0.362  \\
	        & (0.728) & (0.029) & (0.361) \\
          Se    &  0.362  &  0.245  &  0.154  \\
                & (0.358) & (0.235) & (0.149) \\
\end{tabular}
\end{table}
The study of $\alpha$-Ag$_2$Se, with three lattice parameters 
and nine internal parameters to relax, is a stringent test of 
the quality of our pseudopotentials. All the calculations on $\alpha$-Ag$_2$Se
were done with a plane-wave cut-off of 400 eV. This cut-off yields
well-converged properties of the fully relaxed structure, since increasing 
$E_{\rm cut}$ to 600~eV changed the forces on the atoms and the stresses on 
the unit cell only by a negligible amount.  
The Brillouin zone (BZ) sampling was performed using a Monkhorst-Pack 
special $k$-points mesh\cite{monk76}. We also had to introduce a Gaussian 
smearing of the Fermi surface\cite{gil89} due to the absence of a
real gap in the system.  We had to use a (4$\times$4$\times$4) grid (equivalent
to 8 $k$-points in the irreducible part of the BZ) for the energy per atom
to be converged within a few meV. The energy broadening we used was 0.2 eV.

We performed a full structural relaxation of the crystal structure by
varying the three lattice parameters and the nine independent
internal parameters.  In Tab.\ \ref{tab:solid}, the calculated lattice 
constants and internal parameters are compared with experiment. The lattice 
parameters are in satisfactory agreement with experiment, being too low by
$\sim$2\%, which is typical of LDA calculations.
The experimental internal parameters for the Ag atoms are very well 
reproduced. For the Se atoms however, the relaxations were larger, and
this leads to larger discrepancies with experiment. The resulting Ag-Se
distances compared with experimental values 
are (\AA\ units): 2.68 (2.62), 2.72 (2.78), 
2.89 (2.86), 2.80 (2.71) for Ag$_{\mbox{\scriptsize I}}$ and 2.67 (2.74), 
2.74 (2.72) and 2.80 (2.81) for Ag$_{\mbox{\scriptsize II}}$.

\begin{figure}[b]
\begin{picture}(85,70)(5,0)
\psfig{figure=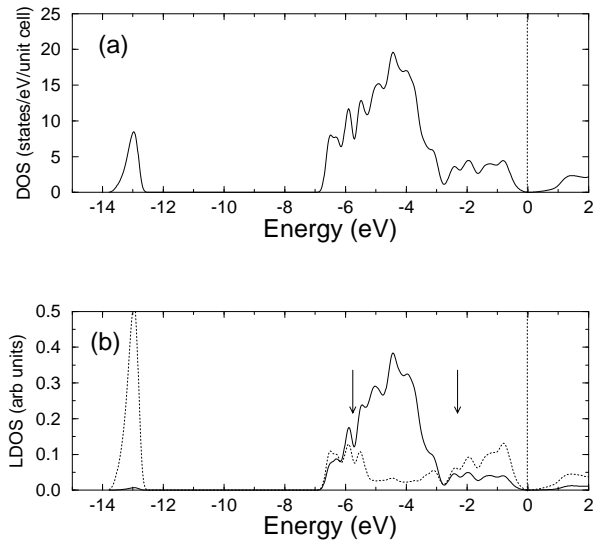,width=8.5cm}
\end{picture}
\caption{(a) Electronic density of states of solid \protect{$\alpha$-Ag$_2$Se}
and (b) local density of states on the Ag (full line) and Se (dotted line) 
sites. The arrows indicate the energies of the states analysed in Fig.\ 3.}
\label{fig:solid_dos}
\end{figure}
In Fig.\ \ref{fig:solid_dos}a, we show the electronic 
density of states (DOS) of 
$\alpha$-Ag$_2$Se. The local densities of states (LDOS) shown in 
Fig.\ \ref{fig:solid_dos}b  
allow us to identify the main features of the DOS. The Se(4$s$) states 
give rise to the lowest peak at --12.5~eV, while the large peak below --4~eV 
originates from the Ag(4$d$) states. This prominent peak is superimposed on 
a broader feature extending from --6.5~eV up to the Fermi 
level, which arises from the Se(4$p$) states. 
Even though the DOS is very small at the Fermi level, there is no real gap
in the system. This is not surprising in view of the very small value
of the experimental gap (less than 0.25 eV) and the well known tendency
of DFT to give energy gaps that are too small. 
An examination of the band structure reveals
that the states at the Fermi level are situated around the $\Gamma$ point in
the BZ.

The LDOS reveals that the Se(4$p$) band is composed of two peaks situated 
at --5.5~eV and --1.5~eV.
These peaks coincide with features in the LDOS on the silver atoms, which
exhibits an asymmetric shoulder below its main peak and has a tail that
extends from --3~eV up to the Fermi level. We believe this to be a
signature of hybridization between the Ag(4$d$) and the
Se(4$p$) states. This hypothesis is clearly supported by 
examining density plots arising from single bands. 
\begin{figure}[b]
\begin{picture}(85,50)(5,5)
\psfig{figure=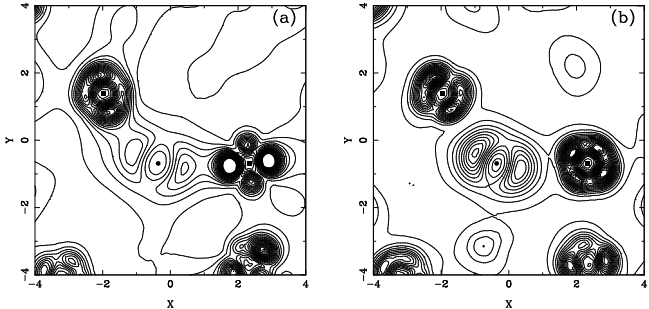,angle=-90,width=9cm}
\end{picture}
\caption{Square of Kohn-Sham orbital from two bands with energies of --5.76 and
--2.30 eV respectively. The orbitals are plotted on a plane
passing through a Se (circle) and two neighboring Ag (square).}
\label{fig:rhopsi}
\end{figure}
In Fig.\ \ref{fig:rhopsi} we show the densities of two states, in
a plane passing through a Se atom and two neighboring Ag atoms.
The energies of the two states are indicated by arrows in 
Fig.\ \ref{fig:solid_dos}. For both energies, the density is 
distributed over both the Se and the Ag atoms.
The wave-function clearly has a $p$ character on the Se, whereas
on the Ag sites the four visible lobes indicate a pronounced
$d$-like character. 

An alternative way of looking at the bonding in a solid is to examine how the
charge is redistributed with respect to neutral atoms. A careful analysis of 
this charge redistribution in $\alpha$-Ag$_2$Se shows that the region where
the electron density has been enhanced is centered on the Se atoms and is 
broadly -- almost spherically -- spread between the Se and its nearest Ag atoms.
This non-directional character of the bonds together with the high
coordination of Se are compatible with ionic bonding.

The picture that emerges is thus a valence band consisting of Se(4$s$), 
Se(4$p$) and Ag (4$d$) states with strong hybridization between the latter two.
The empty states above the Fermi level can be seen as Ag(5$s$/$p$) states,
presumably hybridized quite strongly with Se(4$p$).
With the Se(4$p$) band filled and the Ag(5$s$/$p$) states empty we thus
arrive at a (partially) ionic model for the stoichiometric solid.
\section{Liquid alloys}

\subsection{Details of the simulations}

Our simulations of the Ag-Se liquid alloys have been performed
on a system of 69 atoms in a cubic box with periodic
boundary conditions. The wave-functions are expanded in plane waves
with the same cutoff of 400~eV as before; 
this implies a basis set of $\sim$~28,000
plane waves for the whole system.
We included only the $\Gamma$-point to sample the Brillouin zone, and 
we used a Fermi-smearing energy width of 0.2 eV, as for the solid.
The Verlet algorithm was used to integrate the ionic equation of
motion, with a time step of 3~fs.
We have performed simulations at the temperature $T \simeq$~1350~K 
for three concentrations of Ag$_{1-x}$Se$_{x}$, 
namely: $x=0.33$ (46 Ag atoms and 23 Se atoms), $x=0.42$ (40 Ag atoms and 29
Se atoms) and $x=0.65$ (24 Ag atoms and 45 Se atoms).
The simulations were performed at a density which is linearly interpolated
between experimental values for $\ell$-Ag$_2$Se \cite{glazov70} and 
$\ell$-Se\cite{waseda}.

To initiate the simulations, we exploited the fact that
an empirical pair-potential model \cite{rino88} has been developed for the
stoichiometric Ag$_2$Se system. This is a partially ionic model which 
reproduces the structure of the liquid reasonably when used in classical
molecular dynamics -- a direct comparison of the model with
our AIMD simulations will be presented later. 
We began by making simulations with this
empirical model, and we then switched over to AIMD and let the system
equilibrate for a further 1~ps before collecting data over the
next 2~ps. We reached the other Ag$_{1-x}$Se$_x$ compositions by
replacing some of the Ag atoms by Se atoms and then equilibrating
for 1~ps at the new composition; production runs of 1-2~ps were
again performed in each case.

\subsection{Structural properties}

\subsubsection{Structure factors}

The most direct and detailed way to compare our simulations with
experiment is through the static structure factors, which have been
measured for the stoichiometric composition Ag$_2$Se. The neutron-weighted
structure factor $S_n (k)$ was measured some years ago\cite{price93}, and
 very recently
the partial structure factors $S_{\alpha \beta} (k)$ have also been
measured, using the technique of isotope substitution\cite{barnes96}. 
The definition of $S_{\alpha \beta} (k)$ used here to compare with experiment is
that due to Faber and Ziman\cite{saboungi90}, according to which:
\begin{equation}
 S_{\alpha \beta}(k)=(\langle \rho_{\alpha}^{\ast}(\mbox{\boldmath $k$})\rho_{\beta}(\mbox{\boldmath $k$})\rangle
-\delta_{\alpha \beta})/\sqrt{c_\alpha c_\beta}, 
\end{equation}
where $c_\alpha$ is the concentration of species $\alpha$, 
$\rho_\alpha ( \mbox{\boldmath $k$} )$ represents the Fourier transform of
the number density of species $\alpha$, and the angular brackets
indicate the thermal average. With this definition, the neutron-weighted
structure factor is given by:
\begin{equation}
 S_{n}(k)=\frac{\sum_{\alpha\beta}{c_{\alpha}c_{\beta}}{b_{\alpha}b_{\beta}}S_{\alpha \beta}(k)}{\sum_{\alpha}c_{\alpha}b_{\alpha}^{2}} + 1,
\end{equation}
where $b_\alpha$ is the coherent
scattering length of species $\alpha$. We use the scattering
lengths $b_{\rm Ag}$ = 5.98~fm and $b_{\rm Se}$ = 7.97~fm~\cite{end90,price93}.

\begin{figure}[t]
\begin{picture}(85,70)(12,5)
\psfig{figure=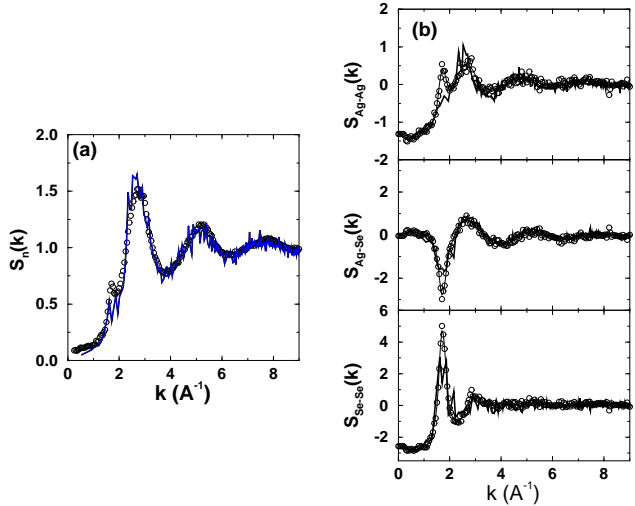,angle=-90,width=10cm}
\end{picture}
\caption{(a) Comparison of the calculated total neutron weighted structure 
factor $S_{n}(k)$ of Ag$_2$Se with the experimental results 
\protect{\cite{end90}}.
(b) The measured partial structure factors S$_{\alpha\beta}(k)$ of
Ag$_2$Se \protect{\cite{barnes96}} compared with those extracted from our 
simulation. The theoretical curves are shown in thick lines, those from
experiment with circles (the thin line is a smoothed version of the experimental
curve).}
\label{sfn}
\end{figure}

Comparisons of the structure factors given by simulation and experiment
are shown in Fig.~\ref{sfn}. Agreement between the two for $S_n (k)$ is very good,
with all the main peaks having the correct position and height. The
excellent agreement of the period, amplitude and phase of the oscillations
at larger wavevectors is particularly gratifying. Our results also
reproduce the small pre-peak at $k \simeq$ 1.7~\AA$^{-1}$,
though its magnitude is somewhat lower than the experimental value.

However, the more detailed comparisons allowed by the $S_{\alpha \beta} (k)$
structure factors reveal some discrepancies. The most noticeable is the
absence of the pre-peak at $k \simeq$ 1.7~\AA$^{-1}$ in our
$S_{\rm Ag-Ag} (k)$. There are also substantial differences in the
magnitudes of the peaks in $S_{\rm Ag-Se} (k)$ and $S_{\rm Se-Se} (k)$
at this wavevector. Since we found a discrepancy in $S_n (k)$ at the
same wavevector, the problem is unlikely to be due to the
experimental analysis of the structure factor into partials. 
Our initial suspicion was that the small size of our system might
be responsible for the discrepancy. However, we have tested
this idea by doing simulations with the empirical model for different
sizes of system, and these tests give no indication of significant
differences between the 69-ion system and much larger systems.
It is possible that fluctuations at the wavevector in question
are very slow and that our simulations are not long enough to
achieve adequate statistics for such long wavelengths.

\subsubsection{Pair correlation functions}
\label{sec:coord}
\begin{figure}[b]
\begin{picture}(85,80)(5,10)
\psfig{figure=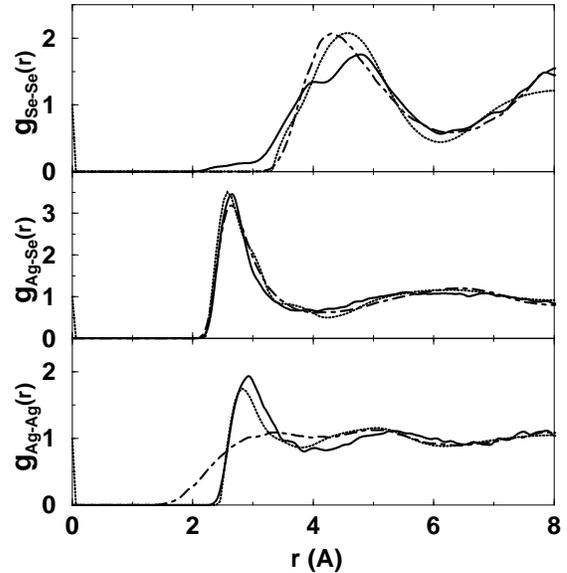,width=8.5cm}
\end{picture}
\caption{Calculated partial radial distribution functions $g_{\alpha\beta}(r)$
of Ag$_2$Se (full lines) compared with the results of experiment \protect{\cite{barnes96}} (dotted lines) and the predictions of an empirical ionic
model \protect{\cite{rino88}} (dash-dotted lines).}
\label{fig:rdf}
\end{figure}

Our calculated partial radial distribution functions (RDF) 
$g_{\rm Ag - Ag} (r)$, $g_{\rm Ag - Se} (r)$ and $g_{\rm Se - Se} (r)$
for the stoichiometric composition are reported in Fig.\ \ref{fig:rdf}, 
where we compare
with the neutron diffraction results of Lague {\em et al.\ }~\cite{barnes96}, 
and with the predictions of the empirical interaction model\cite{rino88}
mentioned earlier.
The agreement of our simulated RDFs with experiment in the region of the first
peak is excellent for Ag-Se and reasonably good for Ag-Ag,
but in both cases there are quite noticeable differences at larger
distances, as would be expected from the discrepancies of
the corresponding $S_{\alpha \beta} (k)$ at $k \simeq$~1.7~\AA$^{-1}$.
The agreement with experiment is somewhat less good for $g_{\rm Se - Se} (r)$,
and there is no sign in the experimental results of the doubling of the first
peak observed in our simulations, or of the weak distribution we
find for $r <$~3~\AA. We return below to the meaning of the latter.
If we are correct in attributing discrepancies in the structure
factors at low $k$ to inadequate sampling, then discrepancies in
the RDFs at large $r$ are presumably due to the same effect. 
However,
the differences of simulated and experimental $g_{\rm Se - Se} (r)$
around the first peak must have a different explanation, and we do not
at present know what this is.

Our comparisons show that the empirical interaction model performs 
extremely well for $g_{\rm Ag - Se} (r)$, but differs greatly 
from both the experimental and AIMD results for $g_{\rm Ag - Ag} (r)$.
It seems certain that the model is seriously mis-representing
the Ag-Ag interaction at short distances. Interestingly, for
$g_{\rm Se - Se} (r)$, the model agrees somewhat better with experiment than 
our AIMD results around the first peak.

\begin{figure}[b]
\begin{picture}(85,60)(10,-18)
\psfig{figure=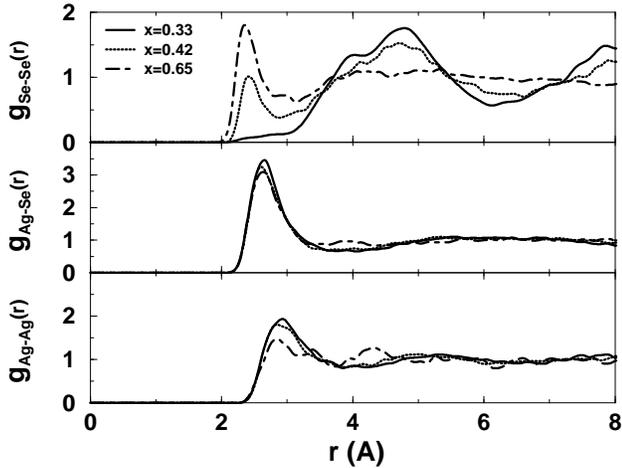,angle=-90,width=8cm}
\end{picture}
\caption{Partial radial distribution functions $g_{\alpha\beta}(r)$ from
simulated $\ell$-Ag$_{1-x}$Se$_x$ at concentrations $x$=0.33 (full line), $x$=0.42 
(dotted line) and $x$=0.65 (dot-dashed line).}
\label{fig:partrdf} 
\end{figure}

We show in Fig.\ \ref{fig:partrdf} how the partial RDFs change with composition
according to the AIMD simulations.
The results show that increase of Se content causes dramatic changes
in $g_{\mbox{\scriptsize Se-Se}}(r)$.
At stoichiometry,
$g_{\mbox{\scriptsize Se-Se}}(r)$ exhibits broad peaks at 3.99 and
4.83~\AA, and only a very weak tail below 3.0~\AA. For $x=0.42$,
instead of a tail, there is a short-distance peak at 2.35~\AA. The
position of the main peak has shifted to 4.72~\AA\ and has decreased
in magnitude. The peak around 4.0~\AA\ seen at stoichiometry has
merged with the main peak to give rise to a shoulder. At the last
concentration, the short-distance peak is dominant and
$g_{\mbox{\scriptsize Se-Se}}(r)$ shows little structure beyond
4.0~\AA , with a low and broad second peak.

The mechanism behind the changes observed in $g_{\mbox{\scriptsize
Se-Se}}(r)$ is the tendency of Se to bond to itself as $x$ exceeds
0.33, and it is no coincidence that the position of the short-distance
peak is almost the same as the Se-Se covalent bond
lengths in crystalline and liquid Se, which are equal to 
2.37~\AA\ \cite{akahama93} and 2.34~\AA\ \cite{edeling81} respectively.  
The nature of this peak can be probed by analysing the coordination of the
selenium atoms.  In general, we define the $\alpha$-$\beta$
coordination as the average number of neighbouring $\beta$ atoms
within a sphere of radius $r_c$ around an $\alpha$ atom.  The growth of
the short-distance peak can be characterized by the Se-Se
coordination, which, for $r_c =$~2.9~\AA , 
we calculate to be 0.1, 0.6 and 1.5 for $x$ =
0.33, 0.42 and 0.65 respectively.  Thus the
coordination grows as $x$ increases, reflecting the tendency of Se
atoms to form bonds. This will become even clearer when we examine
the electronic structure of the liquid below.
In comparison to the modifications observed in $g_{\mbox{\scriptsize
Se-Se}}(r)$, $g_{\mbox{\scriptsize Ag-Se}}(r)$ is hardly affected by
the change of composition. 
The magnitude of the main peak decreases slightly
with increasing $x$ and its position varies only a little: 2.65, 2.60
and 2.64~\AA\ for $x$ = 0.33, 0.42 and 0.65 respectively. At
stoichiometry, a very broad second peak may be distinguished around
$\sim$~6~\AA, but it progressively 
disappears with increasing $x$, and at
$x = 0.65$ $g_{\mbox{\scriptsize Ag-Se}}(r)$ is nearly constant for
$r>$~4.0~\AA.  A similar gradual loss of structure is observed for the
Ag-Ag pair correlation function as the height of the first peak
decreases with increasing Se concentration. This is accompanied by a
slight shift of the first peak of $g_{\mbox{\scriptsize Ag-Ag}}(r)$ to
smaller $r$.  For $x=0.65$ we note the presence of many oscillations
in $g_{\mbox{\scriptsize Ag-Ag}}(r)$. These may partly be due to the
small number of silver atoms at this concentration and the resulting
poorer statistics for the Ag-Ag pair correlation function. However,
the absence of structure in $g_{\mbox{\scriptsize Ag-Ag}}(r)$ at this
concentration clearly indicates that the arrangement of the Ag atoms
is very disordered. Furthermore, the low magnitude of the first peak
suggests that the Ag atoms are in fast exchange with their first
coordination shell, and that the Ag atoms are likely to be very mobile.
The characteristic interatomic distances and coordination numbers
deduced from the partial RDFs are summarized in
Table\ \ref{tab:rdf}, together with those found in both the low
temperature ($\alpha$-Ag$_2$Se) and the superionic ($\beta$-Ag$_2$Se)
phases of Ag$_2$Se. 
\begin{table}[t]
\caption{Inter-atomic distances and coordination numbers in simulated
$\ell$-Ag$_{1-x}$Se$_x$ and in the crystalline phases $\alpha$-Ag$_2$Se and 
$\beta$-Ag$_2$Se. $r_{\alpha\beta}$ is the first neighbor distance between
species $\alpha$ and $\beta$. $r^c_{\alpha\beta}$ is the cut-off radius 
used for calculating the coordination number $n_{\alpha\beta}$.
All distances are given in \AA.}
\setdec 0.0
\label{tab:rdf}
\begin{tabular}{lccccc}
Phase & 
$x=0.33$ & 
$x=0.42$ & 
$x=0.65$ & 
$\alpha$-Ag$_2$Se	     & 
$\beta$-Ag$_2$Se \\
\tableline
$r_{\rm{Ag-Ag}}$ & 2.93      & 2.84      & 2.85 & 2.93      & 3.15 \\
$r_{\rm{Ag-Se}}$ & 2.65      & 2.60      & 2.64 & 2.62      & 2.70 \\
$r_{\rm{Se-Se}}$ & 3.99,4.83 & 2.41,4.72 & 2.35 & 3.98,4.59 & 4.30 \\
\tableline
$r^{c}_{\rm{Ag-Ag}}$ & 3.90     & 3.93 & 3.76 & 3.68 & 4.35 \\
$r^{c}_{\rm{Ag-Se}}$ & 3.77     & 3.65 & 3.62 & 2.90 & 3.45 \\
$r^{c}_{\rm{Se-Se}}$ & 2.9,6.12 & 2.90 & 2.90 & 5.14 & 6.00 \\
\tableline
$n_{\rm{Ag-Ag}}$ & 6.6      &  3.4 & 2.0 & 6 & 10 \\
$n_{\rm{Ag-Se}}$ & 3.8      &  4.0 & 5.1 & 3.5 & 4 \\
$n_{\rm{Se-Se}}$ & 0.1,13.6 &  0.6 & 1.5 & 14 & 14 \\
\end{tabular}
\end{table}
It is interesting to notice that the distances and
coordination found for $x=0.33$ are very close to those found in the
low temperature phase. In particular, the positions of the two
superimposed peaks in $g_{\mbox{\scriptsize Se-Se}}(r)$ are similar to
the two nearest Se-Se distances found in the solid.  This is
surprising, since one would rather expect the melt to resemble
the disordered superionic phase, in which the Ag ions diffuse through
a b.c.c. Se lattice. 

\subsubsection{Analysis of the Se structure}

\begin{figure}[b]
\begin{picture}(85,30)(12,10)
\psfig{figure=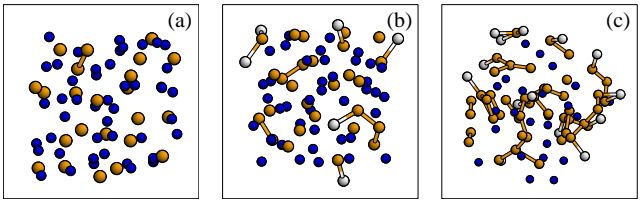,angle=-90,width=8.5cm}
\end{picture}
\caption{Snapshots of typical configurations of $\ell$-Ag$_{1-x}$Se$_x$ at
concentrations (a) $x$=0.33, (b) $x$=0.42 and (c) $x$=0.65. Silver atoms
are shown as black spheres, selenium atoms as gray spheres. Bonds are drawn
between Se atoms with separation $<$ 2.9 \AA. Bonds to atoms in neighboring
cells are represented with two-colored sticks.}
\label{fig:snap} 
\end{figure}

The growth of the short-distance peak in $g_{\rm Se - Se} (r)$ with increasing
Se content is due to the formation of Se clusters for $x >$~0.33. The
simplest way to see this is by studying `ball-and-stick' pictures for typical
configurations. Fig.\ \ref{fig:snap} displays examples of such pictures for the
three compositions. In constructing these pictures, pairs of Se atoms
are joined by sticks if their separation is less than 2.9~\AA, this
distance being chosen because it is close to the first minimum
of  $g_{\rm Se - Se} (r)$ for $x =$~ 0.42 and 0.65. At stoichiometry,
very few Se clusters are present, with only the occasional dimer 
(Fig.\ \ref{fig:snap}a),
and the even more 
\noindent
occasional trimer. It is the presence of these small
clusters that gives rise to the weak tail in $g_{\rm Se-Se} (r)$ at
low $r$. For $x =$~0.42, the Se
atoms bond not only into dimers and trimers but also into larger
Se$_n$ clusters (see Fig.\ \ref{fig:snap}b). At this composition, some
$\sim$~48~\% of the Se atoms are bonded, with 76~\% of these one-fold
and 22~\% two-fold coordinated, with the 
remaining 2~\% having higher
coordination. 
\begin{table}
\caption{Distribution of coordination numbers $n_{\rm{Se-Se}}$ at 
$x$=0.33, 0.42 and 0.65. We show the average percentage 
of atoms with coordination $N_c$.}
\label{tab:secoor}
\begin{tabular}{cccc}
$N_c$ & $x=0.33$ & $x=0.42$ & $x=0.65$ \\
\tableline
0		 &	93  &   52     &  9 \\
1		 &	6   &   36     &  46 \\
2		 &	$<$1   &   10     & 38 \\
3		 &	-   &    $<$2     & 6 \\
$\geq$4	         &	-   &    -     &  $<$1  \\
\end{tabular}
\end{table}
The dominance of one-fold and two-fold coordination
means that most of the clusters are either dimers or Se$_n$ {\em
chains}.  At the composition $x = 0.65$, most of the Se atoms are in
clusters, with an average of 7~\% being isolated (see Fig.\ \ref{fig:snap}c).
The proportions of bonded Se with one-fold and two-fold
coordination are now 35~\% and 40~\%, with a rather significant percentage
(15~\%) of three-fold coordinated atoms. In Table \ref{tab:secoor}, we
summarize the distribution of first-neighbor Se-Se coordination
numbers, for all three concentrations.

\begin{figure}[b]
\begin{picture}(85,75)(5,0)
\psfig{figure=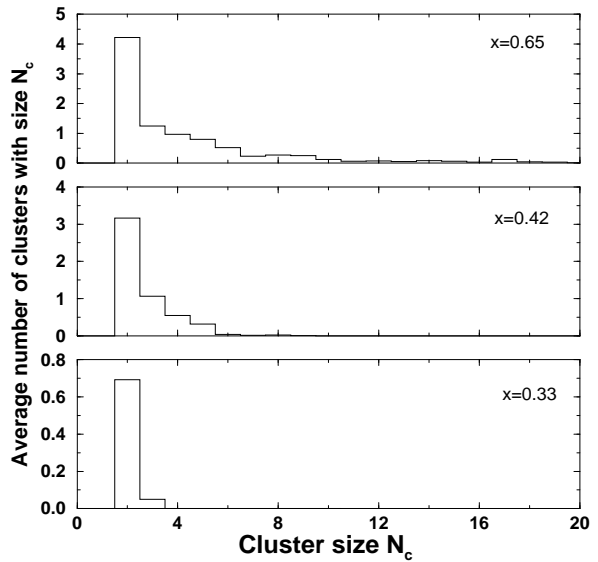,width=8.5cm}
\end{picture}
\caption{Average number of Se clusters of size $n$ in $\ell$-Ag$_{1-x}$Se$_x$ 
at concentrations $x$=0.33, 0.42 and 0.65 (full line).}
\label{fig:clustsize}
\end{figure}

To understand the structure of the Se clusters in more detail, we
have studied their size distribution. In Fig.\ \ref{fig:clustsize}, we
display the average numbers of clusters present in our sample {\em versus}
the number of Se atoms $n$ in the cluster. For all three concentrations, dimers are
the most common clusters. The number of clusters drops rapidly 
with increasing $n$. 
For $x =$~0.42, the
prevalent clusters contain up to 5 atoms. Only a very small number
of clusters with $n >$~5 was found, the largest being $n = 9$.
When the Se content increases to $x = 0.65$, the
cluster-size distribution broadens substantially.  The largest cluster
present in our sample was $n = 28$, but all clusters with
$n < 10$ are very frequent.

An analysis of the structure of the clusters reveals that there
is a simple relationship between the number of bonds within a cluster
$N_b$ and the number of atoms composing that cluster $n$. We found
that for over 95~\% of the clusters we have $N_b =
n-1$, at all three concentrations. The significance of this is that
the relation $N_b = n - 1$ indicates a tree-like topology without closed
loops. Almost all the exceptions to this are clusters with $N_b = n$,
which indicates the existence of a single closed loop. We note
that, for large clusters, loops defined in this way may occur simply as an
artifact of the periodic boundary conditions. In any case, the conclusion
is clear: loops are rare.

\subsection{Dynamical properties}

We have studied the diffusion of atoms in the liquid by calculating
the time-dependent mean square displacement (MSD), defined
in the usual way for species $\alpha$ as:
\begin{equation}
\langle \Delta r_\alpha (t)^2 \rangle = \frac{1}{N_\alpha}
\left\langle \sum_{i=1}^{N_\alpha} | {\bf r}_{\alpha i} ( t + t_0 ) -
{\bf r}_{\alpha i} ( t_0 ) |^2 \right\rangle \; ,
\end{equation}
where the sum goes over all $N_\alpha$ atoms of species $\alpha$,
$t_0$ is an arbitrary time origin, and the angular brackets denote
a thermal average, or equivalently an average over time origins.
For diffusing systems, the MSD is linear in $t$ for large $| t |$,
and the slope is proportional to the self-diffusion 
coefficient $D_\alpha$ of species
$\alpha$:
\begin{equation}
\langle \Delta r_\alpha (t)^2 \rangle \rightarrow 6 D_\alpha | t | +
B_\alpha \; ,
\label{eq:diff}
\end{equation}
where $B_\alpha$ is a constant.

\begin{figure}[b]
\begin{picture}(85,75)(5,0)
\psfig{figure=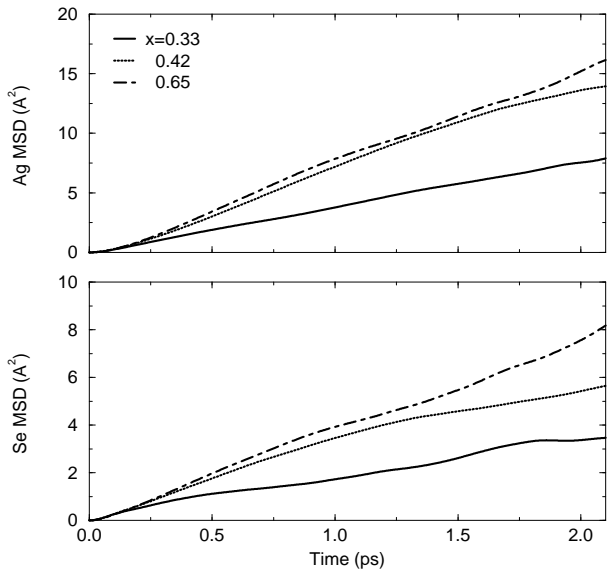,width=8.5cm}
\end{picture}
\caption{Mean square displacement $\langle \Delta r_\alpha (t)^2 \rangle$ of Ag (top panel) and Se (lower panel)
in $\ell$-Ag$_{1-x}$Se$_x$ for $x$=0.33 (full lines), 0.42 (dotted lines)
and 0.65 (dash-dotted lines).}
\label{fig:msd}
\end{figure}
We have calculated the MSD for Ag and Se for the three compositions,
by averaging over the atoms of each species and over time origins.
Our calculated MSDs are graphed in Fig.~\ref{fig:msd}, and the estimated diffusion coefficients are reported in Table \ref{tab:diff}. 
In all normal liquids, the
asymptotic linear behavior of $\langle \Delta r_\alpha (t)^2 \rangle$
is attained after only a few tenths of a ps, and this is the behavior
we find. The lack of straightness of the curves at long times
is simply an effect of statistical averaging (the ideal asymptotic form
shown in Eqn.~(\ref{eq:diff}) assumes perfect averaging over a simulation of infinite
duration). The slopes used to obtain the values of $D_\alpha$ were 
estimated by a least-square fitting of a line to the data.

Our results show three things: the diffusion coefficients are similar to
those of other normal liquids, being in the range 10$^{-5}$ --
10$^{-4}$~cm$^2$s$^{-1}$; Ag diffuses faster than Se; and the diffusion
of both species increases with Se content. Unfortunately, there
appear to be no experimental data for the diffusion coefficients.
However, $D_{\rm Ag}$ has been measured in the superionic phase of
the $\beta$-Ag$_2$Se crystal\cite{okazaki67,mostafa83}. 
This is relevant, because it is commonly found that the diffusion coefficients of the mobile species in
superionic conductors change only a little on melting. The measured
value of $D_{\rm Ag}$ in superionic $\beta$-Ag$_2$Se at the
melting point is $6 \times 10^{-5}$~cm$^2$s$^{-1}$\cite{okazaki67}, which
is rather close to our calculated value of $6.4 \times 10^{-5}$~cm$^2$s$^{-1}$
for the stoichiometric liquid. It is also relevant to note that
the empirical interaction model\cite{rino88} reproduces the experimental
values of $D_{\rm Ag}$ in $\beta$-Ag$_2$Se quite well. This model
predicts the values $D_{\rm Ag} = 10.6 \times 10^{-5}$~cm$^2$s$^{-1}$
and $D_{\rm Se} = 3.4 \times 10^{-5}$cm$^2$s$^{-1}$ for $\ell$-Ag$_2$Se
at $\sim$~1380~K\cite{rino88,clmd}, which are quite close to our AIMD values. 
However, we believe that too much weight should not be put on this
comparison, because of the rather poor results for $g_{\rm Ag - Ag} (r)$
given by the empirical model.

Even though experimental values for the $D_\alpha$ are not available,
there have been measurements of the ionic conductivity $\sigma_i$, i.e. the
electrical conductivity measured under conditions such that the 
flow of electrons is blocked. A rough check against our calculated
$D_\alpha$ can be made using the approximate Nernst-Einstein
relation:
\begin{equation}
\sigma_i = (k_B T)^{-1} \sum_{\alpha} {\bar{\rho}}_\alpha ( z_\alpha e)^2
D_\alpha \; ,
\end{equation}
where ${\bar{\rho}}_\alpha$ is the bulk number density of
species $\alpha$ and $z_\alpha$ is its ionic charge. This relation is
based on the assumption that ions diffuse independently of one another.
The check can only be rough since it is not clear exactly what
values should be assumed for $z_\alpha$. However, if we assume full ionicity 
($z_{\rm Ag} = +1$, $z_{\rm Se} = -2$), then our calculated values of
$D_\alpha$ for the stoichiometric composition lead to a predicted $\sigma_i$ 
of 4.9~$\Omega^{-1}$cm$^{-1}$,
which is rather close to the measured value of 
5.2~$\Omega^{-1}$cm$^{-1}$\cite{endo80,okazaki67,mostafa83}.

\begin{table}
\caption{Self-diffusion coefficients $D_{\alpha}$ of Ag and Se in 
$\ell$-Ag$_{1-x}$Se$_x$ for $x$=0.33, 0.42 and 0.65 in units of 
10$^{-5}$ cm$^2$s$^{-1}$.}
\label{tab:diff}
\begin{tabular}{cccc}
$x$ & 0.33 & 0.42 & 0.65 \\
\tableline
$D_{\mbox{\scriptsize Ag}}$ & 6.4 & 11.2 & 13.1 \\
$D_{\mbox{\scriptsize Se}}$ & 2.7 & 4.2 & 6.3 \\
\end{tabular}
\end{table}
Our finding that $D_{\rm Ag}$ and $D_{\rm Se}$ increase with Se
content is surprising at first sight, since we have seen that Se
forms large clusters when $x >$~0.3. As noted in the Introduction,
the chain-like nature of pure $\ell$-Se is also well established.
It would be expected that the formation of covalently bonded clusters
would make it more difficult for Se atoms to diffuse, and that
the chain-like network would also hinder the diffusion of Ag. However,
this depends very much on the stability of the chains, in other words
on the lifetime of the covalent bonds.

\begin{figure}[b]
\begin{picture}(85,75)(5,0)
\psfig{figure=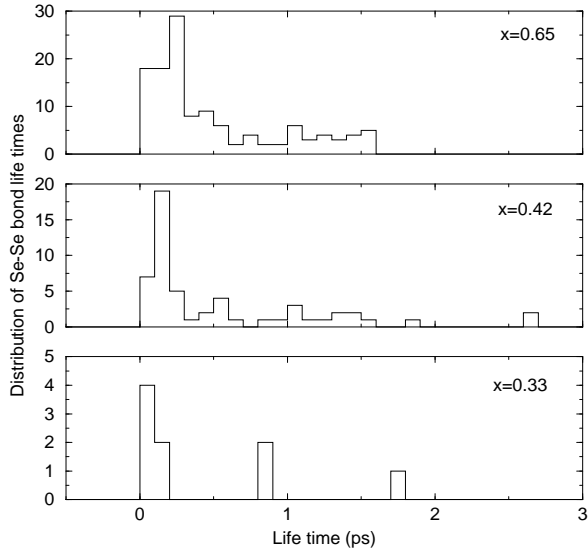,width=8.5cm}
\end{picture}
\caption{Distribution of Se-Se bond lifetimes in $\ell$-Ag$_{1-x}$Se$_x$ for 
$x$=0.33, 0.42 and 0.65.}
\label{fig:life}
\end{figure}
In order to pursue this question further, we have analyzed our simulations
to estimate the Se-Se bond lifetime. To define the lifetime of a bond
between two atoms, we regard the bond as being made when the distance 
between the atoms becomes less than a creation distance $r_1$; it is
broken when the distance becomes greater than a somewhat larger cut-off
$r_2$. We decided to adopt this definition involving two cutoff radii
because there are large fluctuations in the distance between bonded
atoms, and we do not wish to count a bond as broken simply because 
the distance momentarily exceeds the creation distance $r_1$. In fact,
our conclusions would not be significantly affected if we put $r_1 = r_2$.
The distribution of bond lifetimes obtained for $r_1 =$~2.9~\AA\ 
and $r_2 =$~3.2~\AA\ is shown for all three compositions in 
Fig.\ \ref{fig:life}. The distributions are very broad, with lifetimes ranging from less than
0.1~ps up to 2.6~ps. For the majority of bonds, the lifetime is
less than 1.0~ps. For $x =$~0.42 and 0.65, the average lifetimes are
0.52 and 0.50~ps respectively. (The number of Se-Se bonds in the stoichiometric
case was too small to give a statistically meaningful lifetime.)
These rather short bond lifetimes make it less surprising that the diffusion
coefficients increase with increasing Se content.

\subsection{Electronic properties and bonding}
\label{sec:elect}

\begin{figure}[b]
\begin{picture}(85,50)(5,5)
\psfig{figure=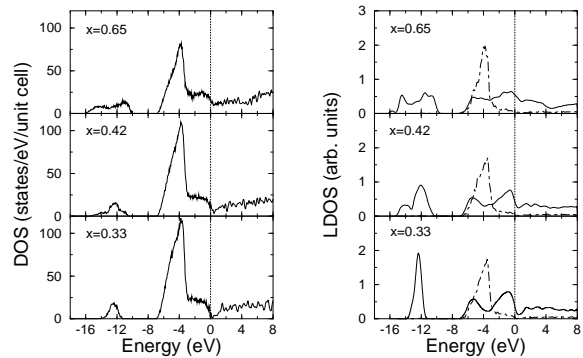,angle=-90,width=9cm}
\end{picture}
\caption{Density of states (left panel) and local densities of
states (right panel) for Ag (chain curve) and Se (dotted curve) from
simulations of Ag$_{1-x}$Se$_x$ at $x$ = 0.33, 0.42 and 0.65.
For clarity, the scale used for the Se LDOS is four times that
used for the Ag LDOS. The vertical dotted line marks the Fermi
energy.}
\label{fig:dos} 
\end{figure}

In this section we will examine how the structural changes are linked to 
the electronic properties of the liquid alloy. 
The calculated electronic DOS for the three compositions are shown in 
Fig. \ref{fig:dos}a. The states can be classified as: Se(4$s$) states 
around --12~eV, Ag(4$d$) states at --4~eV and Se(4$p$) states 
in the region above --7~eV.  The identification of these 
features is made straightforward by the examination of the LDOS 
on the Ag and Se atoms, which are displayed in Fig. \ref{fig:dos}b.

The DOS and LDOS for $\ell$-Ag$_2$Se are very similar to those found for 
$\alpha$-Ag$_2$Se (see Sec.\ \ref{sec:solid}). In particular, the
hybridization of Se(4$p$) states with Ag(4$d$) states is also found
in the liquid, as is clear from the asymmetry of the Ag(4$d$) band and 
the dumbbell-shaped Se(4$p$) band. We note however an increase in the DOS
in the region of the Fermi energy on going from solid to liquid.

Even though the DOS for the three concentrations are quite similar, we 
observe significant modifications as the Se concentration is changed.
First, the height of the sharp Ag(4$d$) peak decreases with respect to
the higher peak of the Se(4$p$) band which remains the same. Second,
the Se(4$s$) band which is narrow and has a single peak at stoichiometry
becomes substantially broader and splits into two peaks as $x$ is increased.
Finally, the DOS at the Fermi level ($E_{\rm F}$) increases with $x$ and
the pseudo-gap present at $x$=0.33 becomes shallower.

The LDOS reveal the nature of these changes. We see that the shape of
the Ag LDOS is essentially identical for all three concentrations. 
In particular, the asymmetry of the main peak is present for all $x$, which 
means that the Ag(4$d$)-Se(4$p$) 
hybridization is unaffected by changes in the 
concentration. The rigidity of the Ag LDOS thus implies that the 
modifications in the DOS are mainly due to the changes in the LDOS of Se.
There are indeed large changes in the Se LDOS as $x$ changes from 0.33
to 0.65: the two peaks of the $p$ band are smeared out and a third peak 
appears above the Fermi level. The first and third peaks are situated
symmetrically about the middle peak and are approximately of the same
magnitude. The structure of the Se LDOS at $x=0.42$ may simply be described
as intermediate between these two cases, as can be seen from the 
Se(4$s$) states.
For all three concentrations, we see that the states at $E_{\rm F}$
arise from the Se(4$p$) band. Thus the increase of 
the DOS at the Fermi level is 
clearly due to the broadening of the second Se(4$p$) peak, and to the presence
of the third peak which both effectively reduce the depth of the pseudo-gap.

\begin{figure}[b]
\begin{picture}(85,75)(5,0)
\psfig{figure=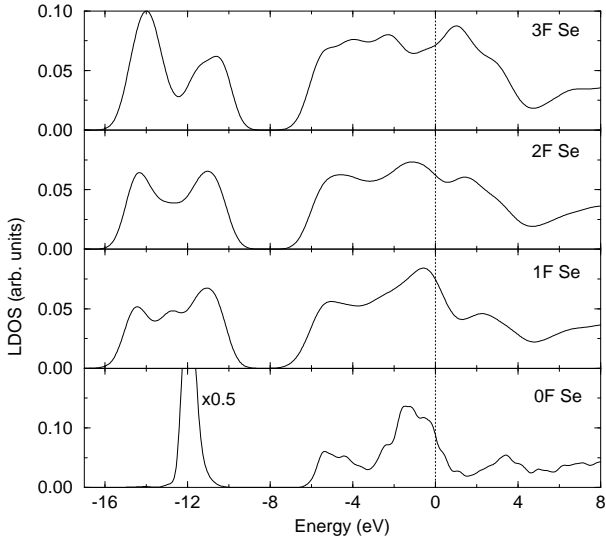,width=8.5cm}
\end{picture}
\caption{Decomposition of the Se LDOS into contributions from isolated (0F),
one-fold (1F), two-fold (2F) and three-fold (3F) coordinated Se atoms.}
\label{fig:ldos_fold}
\end{figure}
All these changes are closely related to the modifications of the
structure. Our interpretation of the changes is as follows.
The presence of a pseudo-gap and the low value of the Se LDOS above
the Fermi level at the stoichiometric composition suggest that all the 
Se(4$p$) states are filled, so that there are 6 $p$-electrons 
per Se atom. This corresponds 
to an ionic model 
in which it is energetically favorable for the Ag(5$s$) electrons to
transfer to empty Se(4$p$) states, since the atomic Ag(5$s$) state is higher 
in energy than the Se(4$p$) states. This corresponds to a charge transfer
from the Ag to the Se ions, thus leading to Ag$^+$ ions 
and Se$^{2-}$ ions. This naturally leads one to interpret the 
states above $E_{\rm F}$ as being Ag(5$s$-$p$) states, so that the picture
here is essentially the same as for the solid, where we have already seen
that an ionic model is appropriate.
As $x$ increases beyond 0.33, the number of occupied 4$p$ states per Se
atom falls below 3 due to the creation of Se-Se bonds. The formation
of these bonds is clearly illustrated by the appearance of the third peak
above the Fermi level, which corresponds to the anti-bonding states, the
first and second peak corresponding to bonding and lone-pair states,
respectively. The anti-bonding states are hybridized with Ag(5$s$-$p$)
states, and only become clearly visible at $x = 0.65$.
The formation of Se-Se bonds is also reflected by the splitting
of the Se(4$s$) which arises from the formation of bonding and
anti-bonding Se(4$s$) combinations in the clusters.

To show more clearly the effect of the formation of Se-Se bonds on the
Se LDOS, we have decomposed this LDOS for $x=0.65$ into contributions
from isolated, one-fold, two-fold and three-fold  coordinated 
Se atoms, as shown in Fig.\ \ref{fig:ldos_fold} (we denote these different
coordinations as 0F, 1F, 2F and 3F). 
We note the strong resemblance of the LDOS for the isolated atoms 
with the Se LDOS for the stoichiometric composition, with the Fermi
close to minimum of the LDOS.
For 2F atoms, the Se(4$s$) band has split symmetrically into bonding and
anti-bonding states, while the Se(4$p$)  now consists of three peaks, which we 
have identified above as bonding, non-bonding (lone pair) and anti-bonding.
The Fermi energy  in this case falls roughly between non-bonding and
anti-bonding states, as we would expect in a chain-like 
structure\cite{robertson79}. The LDOS for 1F atoms is intermediate between the 
0F and 2F forms; the Se(4$p$) bonding and anti-bonding peaks are weaker than
for the 2F atoms, since only one bond is involved.
Note that if $E_{\rm F}$ fell exactly  at the top of the non-bonding peak
the number of electrons in the occupied Se(4$p$) states would give the
0F, 1F and 2F atoms charges of --2, --1 and 0 respectively.


\begin{figure}[b]
\begin{picture}(85,50)(5,5)
\psfig{figure=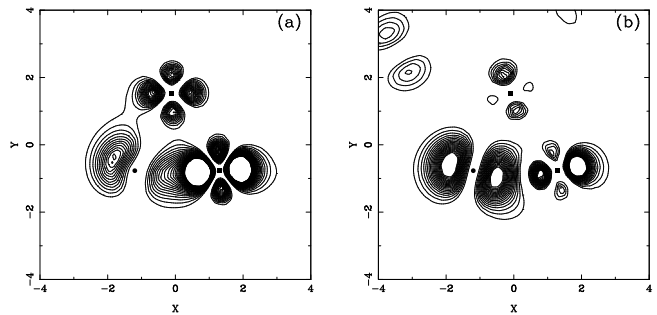,angle=-90,width=9cm}
\end{picture}
\caption{Square of Kohn-Sham orbital from two individual bands with energies 
of --5.3 (a) and --1.0~eV (b) respectively. The orbitals are plotted on a plane 
passing through a Se (circle) and two neighboring Ag (square). 
The hybridized nature of the states is clearly visible, with a $p$ and $d$ 
character of the wave functions on
the Se and Ag atoms respectively. The hybridization has a bonding character
for the lowest states and is anti-bonding for the highest states.} 
\label{fig:rho_hybr}
\end{figure}
\begin{figure}[b]
\begin{picture}(85,60)(5,0)
\psfig{figure=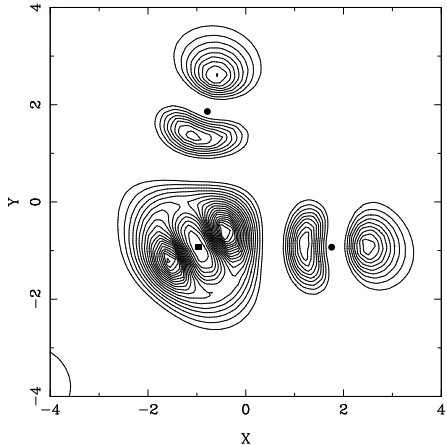,angle=-90,width=9cm}
\end{picture}
\caption{Square of Kohn-Sham orbital of a state from the conduction band
with an energy of ~3.5~eV. The orbitals is plotted on a plane 
passing through a Ag (square) and two neighboring Se (circle).
This state is mainly localised on the Ag atoms
and has a clear $s$ character. This supports the ionic model in which the
silver atoms have lost their $s$ electrons to become Ag$^+$ ions.}
\label{fig:rho_ag_s}
\end{figure}

Several features of the DOS and LDOS can be confirmed by a study of the electron
density. A useful way to do this in the liquid is to examine the density
in a plane passing through three neighbouring atoms.
The presence of hybridization of the Se(4$p$) states
with Ag(4$d$) states is illustrated in Fig.\ \ref{fig:rho_hybr} which shows the
density of one state from the bottom and another from the top of the 
Se(4$p$)/Ag(4$d$) band at --5.3~eV and --1.0~eV, respectively. 
The very close resemblance of these density distributions to those found in the
crystal (see Fig.\ \ref{fig:rhopsi}) shows that this aspect of the electronic
structure persists essentially unchanged in the liquid state. 
In the ionic model we propose the states in
the Ag LDOS above the the Fermi level arise from the empty silver $s$
states. This is confirmed by the charge density of a state at 2~eV above
the Fermi level shown in Fig.\ \ref{fig:rho_ag_s}. The wave function has
a clear $s$ character on the Ag atom and is slightly hybridized with
Se(4$p$) states.


We have calculated the electronic conductivity $\sigma$ in our
simulations by the Kubo-Greenwood formula \cite{mott} as 
has been done in several previous {\em ab initio} calculations 
\cite{galli91,gase96,sti89,dewijs}. 
Our results are shown in Table \ref{tab:conduc} where they are
compared with recent experimental measurements made over the whole 
composition range \cite{ohno94,ohno96}. We have included in the table
a calculated result for the conductivity of $\ell$-Se taken from our own
unpublished AIMD simulation performed at 1370~K \cite{selamun}, since this
will be relevant to the discussion.
\begin{table}
\caption{Calculated conductivity (units of $\Omega^{-1}$cm$^{-1}$) using the 
Kubo-Greenwood formula compared with experimental data \protect\cite{ohno96} at
different compositions $x$.}
\begin{tabular}{ccc}
$x$ & Theory & Expt. \\
\tableline
0.33 & 480  & 410 \\
0.42 & 1500 & 300 \\
0.65 & 1250 & 490 \\
1.00 & 400\tablenote{Ref.\ \cite{selamun}}  & $\sim$1\tablenote{Ref.\ \cite{tamura90}}
\end{tabular}
\label{tab:conduc}
\end{table}
The conductivity $\sigma=480 \Omega^{-1}$cm$^{-1}$ found at 
stoichiometry is in satisfactory agreement with the
experimental data. However, as $x$ increases, the calculated $\sigma$ first 
increases dramatically to $\sim$1500 $\Omega^{-1}$cm$^{-1}$ at $x=0.42$
and then decreases slightly to $\sim$~1250 $\Omega^{-1}$cm$^{-1}$
at $x=0.65$. These values are 3-4 times higher than the experimental
values. For pure $\ell$-Se the calculated $\sigma$ is too large by at least
two orders of magnitude.
We believe that these large discrepancies with experiment arise from the
approximations implicit in the Kubo-Greenwood approach.
There are two major deficiencies in the Kubo-Greenwood formula as applied
here. First, it completely ignores electron-electron collisions, and
includes only the effect of scattering of the electrons by the ions.
Second, and much more serious, it works with the Kohn-Sham energies,
rather than with quasi-particle energies. 
For metallic systems having free-electron-like DOS, the errors are
probably small, and in cases like $\ell$-Si \cite{sti89} and 
$\ell$-Ga\cite{ga95} the method has been shown to work well. But in systems
having a pseudogap, like the liquid alloys considered here, the use of 
the (artificial) Kohn-Sham energies is very likely to cause serious errors.
The case of pure $\ell$-Se is particularly instructive, since experimentally
there is a band-gap of $\sim$0.7 eV \cite{hosokawa90} under the conditions
of interest, whereas our AIMD simulations show no gap, but only a deep minimum
in the DOS. It is important to be clear that this does not point to any error 
in the DFT calculations as far as liquid structure and dynamics are 
concerned. What is at issue here is the (unsurprising) consequence of 
trying to use Kohn-Sham energies in an unjustified way. Similar
large overestimates of the electronic conductivity have been reported
in recent AIMD simulations on $\ell$-Mg$_3$Bi$_2$\cite{dewijs}.
However, it is not yet clear under what circumstances one should
expect these large errors, and it is worth noting that the conductivity
calculated in our very recent AIMD simulations on the $\ell$-Ga-Se 
system\cite{gase96} agreed quite well with experiment in spite of their 
low values and the existence of a deep pseudogap.

\section{Atomic ordering and electronic properties}

We have remarked in a number of places on the ionic nature of the solid
and liquid Ag-Se system, and the implications of this for understanding
the structure. We now try to bring these remarks together into a coherent
picture. The ionic nature of solid $\alpha$-Ag$_2$Se is clear from an
analysis of the DOS, and we have seen that the electronic structure
changes little an melting, so that $\ell$-Ag$_2$Se can be regarded as
consisting approximatively of Ag$^+$ and Se$^{2-}$ ions. In the more Se-rich
liquid we have used the DOS and LDOS to argue that isolated Se ions are in the
Se$^{2-}$ state, while the chains consist of neutral Se atoms terminated
by Se$^{-}$ dangling bonds. This means that the hyperstoichiometric liquid
consists of Ag$^+$, Se$^{2-}$ and (Se$_n$)$^{2-}$ complexes.
This picture can be tested by using simple charge-balance arguments
to predict the average number of Se-Se bonds in the liquid at any
composition, and hence the coordination number of the short distance
peak in $g_{\rm Se-Se}(r)$. If the system consisted entirely of Ag$^+$ 
and Se$^{2-}$ ions, the net charge on a system of $N_{\rm{Ag}}$ and
$N_{\rm{Se}}$ ions would be $N_{\rm{Ag}}-2N_{\rm{Se}}$. If the
clusters are all (Se$_n$)$^{2-}$ then the formation of every Se-Se bond
reduces the charge by two units. For electroneutrality, the number of bonds must
therefore be $N_{\rm{Se}}-\frac{1}{2}N_{\rm{Ag}}$, and the coordination
number of the short distance peak in $g_{\rm Se-Se}(r)$ must be
$2-N_{\rm{Ag}}/N_{\rm{Se}}=3-1/x$. This formula gives predicted values of
0, 0.62 and 1.46 for the Se-Se coordination number at the compositions
$x=0.33$, 0.42 and 0.65, which are quiet close to the values 0.1, 0.6 and
1.5 that we reported in Sec.\ \ref{sec:coord}. This agreement confirms the
picture we propose.

We can interpret the composition dependence of the experimental electrical
conductivity in the light of this picture. For pure $\ell$-Ag, the DOS is
expected to be rather free-electron-like near $E_{\rm F}$. As Se is added,
electrons are drained from the Ag(5$s/p$) band into the Se(4$p$) states
until, at the stoichiometric composition, $E_{\rm F}$ is in a pseudogap
at the top of the Se(4$p$) band. If the DOS remained rigid beyond
this point, $E_{\rm F}$ would move into a region of higher DOS, and
$\sigma$ would increase strongly. Instead of this, Se-Se bonding
starts to occur, and this creates a pseudogap between Se non-bonding
and antibonding states. It is this that maintains $\sigma$ at low values.
Finally, as pure $\ell$-Se is approached, this pseudogap widens, and
$\sigma$ falls to even lower values. As we have seen in the previous section,
our numerical values of $\sigma$ do not reproduce this behavior quantitatively
in the hyperstoichiometric region.

If this picture is correct, it should also be rather general. It is worth
noting that chalcogen pairing has been observed by diffraction experiment
in some equiatomic liquid mixtures, e.g.\ CuSe\cite{cuse} and KTe\cite{kte}.
The conductivity of $\ell$-Cu-Se has been measured over a range of composition
and behaves in a similar way to that found in our simulations, showing a strong
minimum at the composition Cu$_2$Se and then passing through a maximum in the
region of CuSe before falling to the low values associated with $\ell$-Se.
It would clearly be interesting to investigate these and other related
systems by AIMD simulation.
\section{Conclusions}

The AIMD simulations we have presented lead us to the following
conclusions. The reliability and realism of the DFT-pseudopotential
methods we have used are demonstrated by the satisfactory
predictions of the equilibrium structure of the $\alpha$-Ag$_2$Se
crystal a the quite close agreement with neutron diffraction
results for the partial structure factors and RDFs. Nevertheless,
there are significant discrepancies, which need further investigation.
The simulations have shown that a large structural change in the
Ag$_{1-x}$Se$_x$ liquid begins when the Se content exceeds $x = 0.33$,
and this consists of the covalent bonding of Se atoms to form clusters.
These clusters are mainly chain-like, but at $x = 0.65$ there is
a significant fraction of three-fold coordinated Se atoms; the concentration
of rings is extremely small. The
equilibrium fraction of Se present in the form of clusters can be
understood on a simple charge-balance argument based on an ionic
interpretation. In spite of the Se clustering, the diffusion
coefficients of both Ag and Se increase with Se content. This is
not inconsistent with the existence of the clusters, because the Se-Se bonds
turn out to have very short lifetimes of less than 1~ps. The
electronic density of states of liquid Ag$_2$Se closely resembles that
of the solid. With increasing Se content, the main changes in the
density of states are the splitting of the Se(4$s$) into bonding and 
anti-bonding states and the formation of Se(4$p$) bonding and anti-bonding
states associated with Se chains. The calculated electronic conductivity
is in fair agreement with experimental values for the stoichiometric liquid,
but is too large by a factor of 3 -- 4 for higher Se contents. We have
suggested that this error comes partly from the use of Kohn-Sham
energies rather than quasiparticle energies in the conductivity
calculations.

\acknowledgments

The work was done within the U.K. Car-Parrinello
Consortium, which is supported by the High Performance Computing
Initiative. A time allocation on the Cray T3D at EPCC is acknowledged.
Discussions with J. Enderby and A. Barnes, and
technical help from I.~Bush, M.~Payne, A.~Simpson and J.~White
played a key r\^{o}le.
JMH's work is supported by EPSRC grant GR/H67935.
The work used distributed hardware 
provided by EPSRC grants GR/H31783 and GR/J36266.
Financial support from Biosym/MSI is also acknowledged.


\begin {references} 

\bibitem{end90} J. E. Enderby and A. C. Barnes, Rep.\ Prog.\ 
Phys.\ {\bf 53}, 85 (1990).

\bibitem{hensel79} F. Hensel, Adv.\ Phys.\ {\bf 28}, 555 (1979).

\bibitem{schmuzler76} R. W. Schmuzler, H. Hoshino, R. Fisher, and F. Hensel,
Ber.\ Bunsenges.\ Phys.\ Chem.\  {\bf 80}, 107 (1976).

\bibitem{enderby70} J. E. Enderby and E. W. Collings, J. Non-Cryst. Solids {\bf 4}, 161 (1970).

\bibitem{endo80} H. Endo, M. Yao, and K. Ishida, 
J. Phys.\ Soc.\ Japan {\bf 48}, 235 (1980).

\bibitem{glazov86} V. M. Glazov, S. M. Memedov, and A. S. Burkhanov, Sov.\ Phys.-Semicond.\ {\bf 20}, 263 (1986).

\bibitem{ohno94} S. Ohno, A. C. Barnes, and J. E. Enderby, J. Phys.: Condens. Matter {\bf 6}, 5335 (1994).

\bibitem{ohno96} S. Ohno, T. Okada, A. C. Barnes, and J. E. Enderby, 
J. Non-Cryst. Solids, at press.

\bibitem{sigma} In detail the behavior of $\sigma$ in the stoichiometric region
is more complicated. There is a small peak as a function of $x$ at $x$=0.33,
and in a narrow region of composition $\sigma$ anomalously has a negative
temperature derivative (i.e.\ metallic)\protect{\cite{endo80,glazov86,ohno94,ohno96}}. 

\bibitem{tamura90} See e.\ g.\ K. Tamura, J. Non-Cryst. Solids 
{\bf 117/118}, 450 (1990) and references therein.

\bibitem{hosokawa90} S. Hosokawa and K. Tamura, J. Non-Cryst.\ Solids {\bf 117/118}, 489 (1990).

\bibitem{waseda} Y. Waseda, {\em Structure of Non-Crystalline Materials}
(McGraw-Hill, New York, 1980).

\bibitem{price93} D. L. Price, M.-L. Saboungi, S. Susman, K. J. Volin, 
J. E. Enderby, and
A. C. Barnes, J. Phys.: Condens.\ Matter {\bf 5}, 3087 (1993).

\bibitem{barnes96} S. B. Lague, A. C. Barnes, and P. S. Salmon, private 
communication.

\bibitem{bellissent80} R. Bellissent and G. Tourand, J. Non-Cryst.\ Solids 
{\bf 36/36}, 1221 (1980).

\bibitem{edeling81} M. Edeling and W. Freyland, Ber.\ Bunsenges.\ Phys.
Chem. {\bf 85} (1981) 1049.

\bibitem{tamura92} K. Tamura and S. Hosokawa, Ber.\ Bunsenges.\ Phys.\ Chem.\
{\bf 96} (1992) 681.

\bibitem{hohl91} D. Hohl and R. O. Jones, Phys. Rev. B {\bf 43}, (1991)
3856.

\bibitem{selam} F. Kirchhoff, M. J. Gillan, and J. M. Holender,
J. Non-Cryst. Solids, at press.

\bibitem{cp85} R. Car and M. Parrinello, Phys.\ Rev.\ Lett.\ {\bf 63}, 2471 (1985).

\bibitem{df} See e.g.\ G.  P.  Srivastava and D. Weaire, Adv.\ Phys.\ {\bf 36}, 
463 (1987); J. Ihm, Rep.\ Prog.\ Phys.\ {\bf 51}, 105 (1988); M.  J.  Gillan, 
in {\em Computer Simulation in Materials Science}, eds.  M.  Meyer and V.  Pontikis, p.  257 
(Kluwer, Dordrecht, 1991);
G. Galli and M. Parrinello, ibid, p. 283.

\bibitem{rino88} J. P. Rino, Y. M. M. Hornos, G. A. Antonio, 
I. Ebbsj\"o, R. K. Kalia,
and P. Vashishta, J. Chem.\ Phys.\ {\bf 89}, 7542 (1988).


\bibitem{galli91} G. Galli and M. Parrinello, J. Chem.\ Phys.\ {\bf 95}, 7504 (1991).

\bibitem{schone95} M. Sch\"one, R. Kaschner, and G. Seifert, J. Phys.: 
Condens.  Matter\ {\bf 7}, L19 (1995).

\bibitem{wijs94} G. A. de Wijs, G. Pastore, A. Selloni, and W. van der Lugt,
Europhys. Lett.\ {\bf 27}, 667 (1994).

\bibitem{gase96} J. M. Holender and M. J. Gillan, Phys.\ Rev.\ B, in press.

\bibitem{letter} F. Kirchhoff, J. M. Holender, and M. J. Gillan, 
Europhys.\ Lett., in press.

\bibitem{hohenberg64} P. Hohenberg and W. Kohn, Phys.\ Rev.\ {\bf 136},
B864 (1964).

\bibitem{kohn65} W. Kohn and L. J. Sham, Phys.\ Rev.\ {\bf 140}, A1133 (1965).

\bibitem{jones89} R. O. Jones and O. Gunnarsson, Rev.\ Mod.\ Phys.\ 
{\bf 61}, 689 (1989).

\bibitem{hsc} D. R. Hamann, M. Schl\"uter, and C. Chiang, Phys.\ Rev.\
Lett.\ {\bf 43}, 1494 (1979).

\bibitem{bhs} G. B. Bachelet, D. R. Hamann, and M. Schl\"uter,
Phys.\ Rev.\ B {\bf 26}, 4199 (1982).

\bibitem{gil89} M. J. Gillan, J. Phys.: Condens.  Matter\ {\bf 1}, 689 (1989).

\bibitem{sti89cg} I. \v{S}tich, R. Car, and M.  Parrinello, Phys.\ Rev.\ B{\bf 39}, 4997 (1989).

\bibitem{pay92} M.  C.  Payne, M.  P.  Teter, D.  C.  Allan, T.  A.  Arias, and
J.  D.  Joannopoulos, Rev.\ Mod.\ Phys.\ {\bf 64}, 1045 (1992).  

\bibitem{kre94} G. Kresse and J. Hafner, Phys. Rev.  B\ {\bf 49}, 14251 (1994).

\bibitem{gru94} M.  P.  Grumbach, D. Hohl, R.  M.  Martin, and R.  Car,
J. Phys.: Condens.  Matter\ {\bf 6}, 1999 (1994).  

\bibitem{ga95} J. M. Holender, M. J. Gillan, M. C. Payne, and A. D.
Simpson, Phys. Rev. B {\bf 52}, 967 (1995).

\bibitem{pas92} A. Pasquarello, K. Laasonen, R. Car, C. Lee, and D. Vanderbilt,
Phys.\ Rev.\ Lett.\ {\bf 69}, 1982 (1992).

\bibitem{kre93} G. Kresse and J. Hafner, Phys.\ Rev.\  B {\bf 48}, 13115 (1993).

\bibitem{cetep} L. J. Clarke, I. \v{S}tich, and M. C. Payne,
Comp. Phys. Comm. {\bf 72}, 14 (1992).



\bibitem{lin}J.-S. Lin, A.  Qteish, M.  C.  Payne, and V.  Heine, Phys.\ Rev.\ 
B {\bf 47}, 4174 (1993).  

\bibitem{rap}A.  M.  Rappe, K.  M.  Rabe, E. Kaxiras, and J.  D.  Joannopoulos, 
Phys.\ Rev.\ B {\bf 41}, 1227 (1990).  

\bibitem{ker} G.  P.  Kerker, J. Phys. C\ {\bf 13}, L189 (1980). 

\bibitem{kb} L. Kleinman and D. M. Bylander, Phys.  Rev.  Lett.\ {\bf
48}, 1425 (1982).  

\bibitem{kin91} R. D. King-Smith, M. C. Payne, and J. S.
Lin, Phys.  Rev.  B\ {\bf 44}, 13063 (1991).  


\bibitem{agcl} F. Kirchhoff, J. M. Holender, and M. J. Gillan, Phys.\ Rev.\ B{\bf 49}, 17420 (1994).

\bibitem{rahl36} P. Rahlfs, Z. Phys.\ Chem.\ Abt. B {\bf 31}, 157 (1936).

\bibitem{gmelinag} {\em Gmelin Handbook of Inorganic Chemistry, Silver},
edited by G. Czack, D. Koschel and H. H. Kugier (Springer Verlag, Berlin 1983), Suppl.\ Vol.\ B3, pp.\ 152-187.

\bibitem{saku77} T. Sakuma, K. Iida, K. Honma, and H. Okazaki, J. Phys.\ Soc.\
Japan {\bf 43}, 538 (1977).

\bibitem{cava77} R. J. Cava, F. Reidinger, and B. J. Wuensch, Solid State Commun.\ {\bf 24}, 411 (1977).

\bibitem{cava80} R. J. Cava and D. B. McWhan, Phys.\ Rev.\ Lett.\ {\bf 45},
2046 (1980).

\bibitem{wieg71} G. A. Wiegers, Amer. Mineral. {\bf 56}, 1882 (1971).

\bibitem{wells} A. F. Wells, {\em Structural Inorganic Chemistry},
5th edition, (Clarendon Press, Oxford, 1984).

\bibitem{wyckoff} R.  W.  G.  Wyckoff, {\em Crystal Structures},
2nd edition, vol.\ 1 (Interscience, New York, 1964).

\bibitem{engel68} P. Engel and W. Nowacki, Acta.\ Cryst.\ B {\bf 24}, 77, (1968).

\bibitem{orgel58} L. E. Orgel, J.\ Chem.\ Soc., 4186 (1958).

\bibitem{burd92} J. K. Burdett and O. Eisenstein, Inorg. Chem. {\bf 31}, 1758 (1992).

\bibitem{monk76} H. J. Monkhorst and J. D. Pack, Phys.\ Rev.\ B {\bf 13},
5188 (1976).

\bibitem{glazov70} V. M. Glazov and N. M. Makhmudova, Inorg. Materials (USSR)
{\bf 6}, 1240 (1970) as reproduced in Ref.\ \cite{gmelinag} p.\ 169.

\bibitem{saboungi90} M.-L. Saboungi, W. Geertsma and D. L. Price, Annu.\
Rev.\ Phys.\ Chem.\ {\bf 41}, 207 (1990).

\bibitem{akahama93} Y. Akahama, M. Kobayashi and H. Kawamura,
Phys.\ Rev.\ B{\bf 47}, 20 (1993).

\bibitem{okazaki67} H. Okazaki, J. Phys.\ Soc.\ Japan {\bf 23}, 355 (1967).

\bibitem{mostafa83} S. N. Mostafa, Z. Metalkde.\ {\bf 74}, 188 (1983).

\bibitem{clmd} F. Kirchhoff and M. J. Gillan, unpublished.

\bibitem{robertson79} J. Robertson, Phil.\ Mag.\ B {\bf 6}, 479 (1979).
 
\bibitem{mott} N. F. Mott and E. A. Davies, {\it Electronic Processes
in Non-Crystalline Materials}, (Clarendon, Oxford, 1979).

\bibitem{sti89} I. \v{S}tich, R. Car, and M.  Parrinello, Phys.\ Rev.\ Lett.\ {\bf 63}, 2240 (1989).


\bibitem{dewijs} G. A. de Wijs, PhD Thesis, University of Groningen, 1995.


\bibitem{selamun} F. Kirchhoff, M. J. Gillan, and J. M. Holender, unpublished.



\bibitem{cuse} A. C. Barnes and J. E. Enderby, Phil.\ Mag.\ B {\bf 58}, 497 (1988).

\bibitem{kte}  J. Fortner, M.-L. Saboungi, and 
J. E. Enderby, Phys.\ Rev.\ Lett.\ {\bf 69}, 1415 (1992).

\end{references}
\end{document}